\documentstyle[psfig,conf_iap]{article}

\def\cwcplotmacro#1#2#3#4#5#6#7{\centering \leavevmode
    \vbox to#2{\rule{0pt}{#2}}
    \includegraphics{#1}}

\begin{document}

\heading{Mg II Detection in the Ly$\alpha$ Forest: 
Metal Rich Cloud or LSB Dwarf Galaxy?} 

\par\medskip\noindent

\author{Christopher W. Churchill$^{1}$ and Vincent Le Brun$^{2}$}

\address{Department of Astronomy, Pennsylvania State University, 
University Park, 16802}
\address{Laboratoire d' Astronomie Spatiale du CNRS, B.P. 8,
F--13376, Marseille Cedex 12}
\begin{abstract}
We report the detection of Mg~{\sc ii} absorption in two Ly$\alpha$ forest 
clouds along the line of sight to PKS $0454+039$.  These clouds, at
redshifts $z=0.6248$ and $z=0.9315$, respectively, also have strong
Fe~{\sc ii} absorption.
We present results on the $z=0.9315$ cloud.
The upper limit on $\log N$(H~{\sc i}) is 16.5~cm$^{-2}$.  
Based upon UV background photoionization models, we infer that the
cloud has super--solar metallicity.
Photoionization by a late--type stellar population is ruled out.
\end{abstract}

\section{Scientific Considerations}

Metal--line absorption in the intergalactic medium (IGM,
i.e.~Ly$\alpha$ forest clouds) is astrophysically interesting
because the absorption properties can be exploited to reveal the
chemical enrichment and ionization histories of the universe.
At all redshifts, forest clouds far from bright galaxies may sample
IGM chemical enrichment from the first generation of stars.   
At low redshifts, they may trace a population of low surface 
brightness and/or dwarf galaxies.

Cowie et~al.~\cite{cowie} detected high ionization
C~{\sc iv} and Si~{\sc iv} in the forest at $z\sim 2.5$,
where the UV background (UVB) is likely dominated by ``hard''
radiation from quasars and active galactic nuclei.  
At $z<1$, the UVB intensity is reduced by a factor of $\sim 5$
and its shape may be softened by bright field galaxies.
In short, the IGM ionization conditions may have evolved so that low
ionization species, i.e.~Mg~{\sc ii} and Fe~{\sc ii}, are detectable 
in ``forest'' clouds.  

\section{An LSB Galaxy or Metal Enhanced Cloud?}

We searched a HIRES/Keck spectrum of PKS $0454+039$ for Mg~{\sc ii}
absorption in the Ly$\alpha$ clouds reported by Boiss\'e
et~al.~\cite{boisse}.  The success rate was 2/28, or $\sim 7$\%.
The upper limits are $\log N(\hbox{Mg~{\sc ii}}) =
10.55$--11.06~cm$^{-2}$ for $ 0.11 < W_{r}(1215)<10.1$~{\AA} H~{\sc
i} clouds.
The two discovered systems, at $z$ 0.6248 and 0.9315, have
$W_{r}(1215) = 0.33$ and $0.15$~{\AA}, respectively. 
In Figure 1, the 0.93 system is presented.
The unresolved metal lines were fitted with Voigt profiles and
extensive Monte Carlo simulations were performed to obtain the 
column densities and $b$ parameters.  
We obtained $12.3 < \log N < 12.6$~cm$^{-2}$ for both Mg~{\sc ii} and
Fe~{\sc ii}, with $1.4 < b < 2.2$ km~s$^{-1}$. 
We assumed the cloud is thermally broadened and used the curve of
growth to estimate $14.2 < \log N(\hbox{H~{\sc i}}) < 16.5$~cm$^{-2}$.
We then used CLOUDY \cite{ferland}, once with a $\nu ^{-1.5}$ UV
continuum slope and once with $T\approx 5000$~K blackbody spectrum
[cool stars are suggested by $N(\hbox{Fe~{\sc ii}})/N(\hbox{Mg~{\sc
ii}}) \sim 1$], to explore the cloud chemical and ionization
conditions.  

For the power--law UVB with $J_{-21} = 0.2$, the cloud properties are
$n_H \sim 0.05$~cm$^{-3}$, $D \sim 5$~pc, and $M \sim
10^{-3}$~M$_{\odot}$, with ionization parameter $\log U \sim
-3.5$.  This cloud is inferred to have super--solar metallicity, $0.5 <
[Z/Z_{\odot}] < 1.5$!
In a F702W WFPC/HST image of the $0454+039$ field \cite{lebrun} there
are a few diffuse extended LSB galaxies over the impact parameter
range $20 < Dh^{-1} < 60$ kpc, if they have $z=0.93$.
However, it is difficult to understand how extended super--solar gas
could arise from such diffuse galaxies.

For the blackbody ``cool stars'' UV continuum, the inferred cloud
properties are $n_H \sim 0.1$~cm$^{-3}$, $D \sim 100$~pc, and $M
\sim 600$~M$_{\odot}$, with $\log U \sim +1$.  
The inferred metallicity is $[Z/Z_{\odot}] \sim - 1.2$.
This scenario models a dwarf--like galaxy with a late--type stellar
population. 
The value of $U$ implies an extreme number of stars; this
scenario is astrophysically implausible \cite{cwc}.
We thus conclude that the absorber is metal rich and ionized by the
UVB.

It may be that systems like the one presented here are fairly 
common and comprise a yet--to--be explored class of absorber. 
A sensitive search for Mg~{\sc ii} systems in the forests along the
lines of sight to other QSOs is in progress.

\begin{figure}
\cwcplotmacro{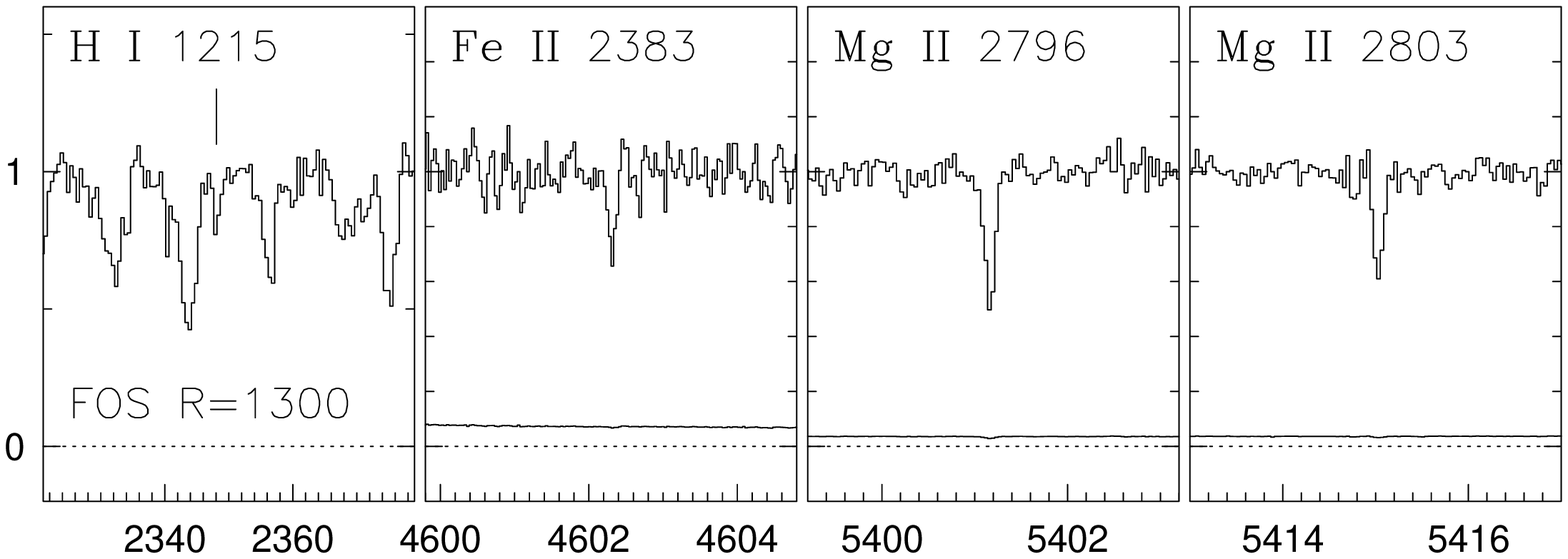}{3.5in}{0}{53.}{45.}{-214}{19}
\vglue -2.29in
\caption[]{The FOS/HST and HIRES/Keck (metal lines) spectra of the
$z=0.93$ cloud.  The tick marks the Ly$\alpha$ $\lambda$1215
absorption line. Only upper limits are found for C~{\sc ii}, C~{\sc
iii}, C~{\sc iv}, Si~{\sc iv}, N~{\sc v}, and O~{\sc vi}.}
\end{figure}

\acknowledgements{The National Science Foundation has supported this
work through AST--9617185.  Thanks to P.~Boiss\'e for providing his
FOS spectrum prior to publication.  Thanks also to J. Bergeron,
J. Charlton, P. Petitjean, and S. Vogt.}

\vfill
\end{document}